\documentclass[fleqn,usenatbib]{mnras}

\usepackage[T1]{fontenc}

\usepackage{soul}
\DeclareRobustCommand{\VAN}[3]{#2}
\let\VANthebibliography\thebibliography
\def\thebibliography{\DeclareRobustCommand{\VAN}[3]{##3}\VANthebibliography}

\usepackage{cite}      
\usepackage{graphicx}
\usepackage{eufrak}

\title[Dark Matter Makes a Difference]{When Leaving the Solar System: Dark Matter Makes a Difference
}

 \author [E.A. Belbruno et al.]{
Edward  Belbruno, $^{1,2}$\thanks{edward.belbruno@yu.edu}
James Green$^{3}$ 
\\
$^{1}$Yeshiva University, Dept. of Mathematics, 500 W 185 St,   New York, New York 10033 USA\\
$^{2}$Princeton University, Dept.  of Astrophysical Sciences,  Ivy Lane, Princeton, NJ 08542 USA \\
$^{3}$NASA HQ, Washington, D.C. 20546-0001 USA
} 

 \date{Accepted 2021 December 21. Received 2021 December 17; in original form 2021 September 13.}
 
 \pubyear{2021}
 
\begin{document}

\label{firstpage}
 \pagerange{\pageref{firstpage}--\pageref{lastpage}}
\maketitle

\begin{abstract}
A resultant gravitational force due to the current estimates of the virial mass of the Milky Way galaxy, dominated by dark matter,  is estimated near the Sun  and is described in two different analytical models yielding consistent results. One is a two step Hernquist model, the other is  a Navarro-Frenk-White  model. The effect of this force is estimated on trajectories for spacecraft sufficiently far from the Sun. The difficulty of detecting this force is studied.  It is concluded that its effect should be considered for certain spacecraft missions. Its effect on the Pioneer and New Horizons spacecrafts is discussed.  A future mission is discussed that may be able to detect this force.   Implications of this force are discussed with its impact for problems in planetary  astronomy and astrophysics.
\end{abstract}





\maketitle

\section{Introduction,  Methodology, Results}  \label{sec:Intro}

When considering the motion of an object in our solar system about the Sun, for example, an asteroid or comet or spacecraft,  the Sun\rq{}s gravity plays a primary role.  If the object is moving far from the Sun, say beyond the orbit of Pluto, the  gravity due to the Solar System (Sun, planets, main belt asteroids, etc. )  is the main gravitational force.  When modeling the motion of a comet or spacecraft or a body such as a Kuiper belt object,  the gravitational force of interest is the Solar System\rq{}s.  The motion of the body is considered to be that of a two-body problem between the body and the centre of mass of the Solar System. 

However, as we show in this paper, if an object is moving sufficiently far from the Sun, then there is another gravitational force that can play an important role. This is the resultant gravitational force of the Milky Way galaxy (MW) and it is primarily generated by dark matter.   We show how to estimate it for our analysis near\footnote{Here {\it near} is a relative term, used in the context of the Milky Way.} the Sun.  Although it is small, it can cumulatively add up and significantly affect the trajectory of motion over long periods of time.

\begin{figure}
\centering
          \includegraphics[width=\columnwidth]{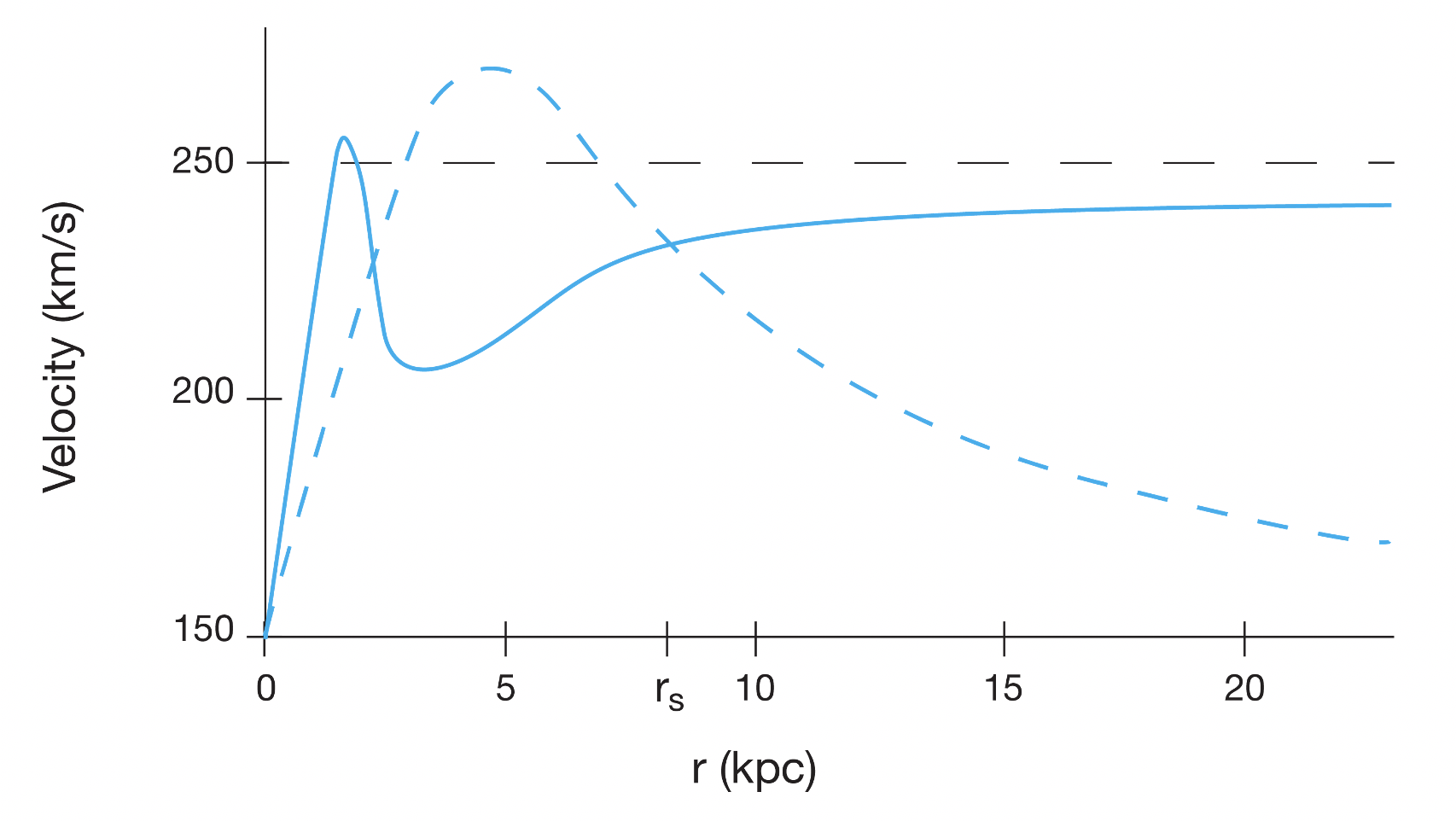}
	\caption{Velocity rotation curves for the Milky Way. The observed one is shown as a solid blue curve reflecting the existence of dark matter and the one that would be predicted by a Keplerian model is the dotted blue curve (\citet{Eilers2019}). This is a rough sketch.}
	\label{fig:fig1}
\end{figure}

It is theorized that in the current model of the observable Universe, ordinary matter (baryonic) consists of $5\%$ of the total energy. Dark matter (non-baryonic), that we cannot see, consists of $25\%$ (\citet{Wechsler2018}).  Neutrinos and photons make up a tiny amount of the energy.\footnote{Dark energy, repulsive in nature, makes up  $70\%$ of the total energy.}  It is estimated  that all but a few percent of the mass of the Milky Way galaxy (MW)  consists of dark matter (\citet{Watkins2019}). 
          The components of the MW consist of the disc, a central bulge (roughly spheroidal and confined to the inner few kpc from the Galactic Centre (GC)),  a stellar halo (spheroidal, extending out 10s of kpc from GC), and the dark matter halo extending out several hundred kpc from GC (\citet{Eilers2019}, \citet{Piffl2014}).
         The dark matter halo contains most of the mass of the Galaxy.   In the Milky Way the dark matter is measured  by observing the rotational  circular motion of the Galaxy about GC and measuring this velocity as a function of radial distance. It is observed that this velocity does not decrease in distance as one would expect from Keplerian motion based on the Newtonian gravitational  inverse square force. Instead,  it  levels off.  This has been measured precisely by numerous observational studies. The predominant conclusion is that this is caused by the dark matter  (\citet{Arbey2021}, \citet{Wechsler2018}).
This is illustrated in Figure \ref{fig:fig1}.

The verification of the existence of dark matter is not just from observing the circular velocities of objects in our Galaxy about the GC, but rather it can indirectly be measured by other independent means in all other galaxies and clusters of galaxies. A way this is done is by using Einstein\rq{}s general theory of relativity to measure the deflection of light as it passes through or near a galaxy, and observing its deflection, called microlensing.  From the angle of the deflection, an estimate can be made of the total mass. The relative  percentage  of dark matter vs  baryonic matter can be obtained by comparing to the estimated mass of bayrons (\citet{Wechsler2018}).   One can estimate the relative fraction of baryons from big bang nucleosynthesis   (\citet{Arbey2021}). 
Another independent measurement is obtained from observing microwave background radiation maps of the early Universe made with the {\em WMAP} and {\em PLANCK} missions showing not only dark matter but also dark energy as well (\citet{Bennett2003}, \citet{Spergel2003}, \citet{Aghanim2020}).   It is remarked that another less common theory exists to explain the deviations of circular motions in the Milky Way that doesn\rq{}t evoke the existence of dark matter called the MOND (Modified Newtonian Gravity)  theory. It hypothesizes that the gravitational field of the Galaxy is non-Newtonian at large distances from GC.  Its accuracy is in question in certain situations where it fails, such as in the Bullet cluster which clearly shows a separation of baryonic and dark matter, and it isn\rq{}t applicable to the early Universe revealed by the microwave background radiation (\citet{Arbey2021}). 

In this paper we will calculate the force per unit mass, or equivalently, acceleration,\footnote{By{\it force} we always mean {\it force per unit mass}. The words force and acceleration are used interchangeably.} on a particle  due to  the dark and baryonic matter of our Galaxy (MW) near the Sun using two different models. This force  is viewed as a perturbation of the Sun\rq{}s gravitational force on the particle.  We are using two different models to show consistent results.  It is first necessary to have a total virial mass of the Galaxy within a galactic halo about GC that extends out several hundred kpc. This is labeled $M_{vir}$. We are using a nominal value  obtained by Watkins et. al. (\citet{Watkins2019}) which is, \footnote{It is remarked that if only baryonic matter were considered, this value would be on the order of $10^{10}$ solar masses, $ M_{\odot}$.} 

\begin{equation}
M_{vir} = 1.54_{-.44}^{+.75} \times   10^{12} M_{\odot} .
\label{eq:GalaxyMass}
\end{equation}

The models considered assume a mass-density as a function of the radial distance $r$ from GC. Within a given spherical region and assuming Newtonian gravity, this mass-density yields a force, $\bf{F_G}$, in GC-centred coordinates. A test for the validity of using these models  is to ensure that ${\bf F_G}$ yields the correct velocity for the Sun about GC which is estimated to lie between $220-250$ km/s (\citet{Watkins2019}).  It is verified in Section \ref{sec:Models} that this is the case.   Moreover, we show that both models yield the exact same value of the force, even though the models are different. 

 The first model we consider in Section \ref{sec:Models} is  a modified Hernquist model.  It has a density profile $\nu(r)$. This model is applied in two steps. The first step applies the Hernquist model to the large component of MW mainly consisting of the dark matter halo and using  (\ref{eq:GalaxyMass}) estimates the mass of the Galaxy, consisting of dark matter as well as baryonic matter  within the distance $r_s$ from GC, where $r_s = 8.29$ kpc is the distance from the Sun to GC,  and in the second step the Hernquist model is applied again for the spherical region within $r_s$, denoted by $\leq r_s$, which is dominated by the smaller stellar halo of baryonic matter.  Then, within this smaller spherical region,  the force near the Sun due to dark  and  baryonic matter is computed.  Two different scale lengths are used, for the two steps.  This model accurately determines the Sun\rq{}s velocity.
 
A second model is also considered as a check to the first model for consistent results.  This is the Navarro-Frenk-White (NFW) model (\citet{Navarro1996}) together with a point-mass Newtonian potential model. The NFW model is the standard model for estimating the dark matter halo. It assumes a mass-density profile $\rho(r)$ to estimate the density of dark matter within a given radial distance $r$ from GC, but the dark matter is predominately in the galactic halo regions and not as concentrated in the solar neighbourhood. From this density variation, the mass of dark matter within this radial distance, $M(r)$,  can be computed.   The force due to dark matter is determined near the Sun, in Section \ref{sec:Models}. In that section it is seen that this model does not accurately determine the Sun\rq{}s velocity about GC since this model is designed to estimate dark matter, and doesn\rq{}t reflect baryonic matter in its modeling, which is relatively more concentrated near the Sun\rq{}s distance from GC. To compensate for this, this model is slightly modified by adding a baryonic mass component using a Newtonian point-mass potential.

 As is shown in Section \ref{sec:Models}, these two models yield the identical force value at the Sun which is  in magnitude, approximately,
 \begin{equation}
F_{G}(r_s) =  1.8 \times 10^{-10} m/s^2  ,
\label{eq:FGValue}
 \end{equation}
 in GC-centred coordinates.  It turns out this force is a central force field with a direction towards GC. Its magnitude only depends on its distance from GC.   Its magnitude at another vector point ${\bf x}$ near the Sun, at a distance $r$ from GC,  is $F_G(r)$. In Sun-centred coordinates this force is calculated as a tidal force  ${\bf F_{G,S}}({\bf x}) =  F_G({\bf x})- F_G({\bf x_s})$, where ${\bf x_s}$ is the vector location of the Sun, at the distance $r_s$ to GC.   This is what would actually be detected on a spacecraft near the Sun. It is relatively small for objects near the Sun and would have to be measured to high precision since it is on the order of 10 million times less than $F_G$.   This is described in Section \ref{sec:Test}  (see (\ref{eq:TidalEqu4})). It is noted that ${\bf F_{G,S}}$ defines an acceleration field at each point ${\bf x}$ near the Sun. It can be defined for a spacecraft regarding it as a point mass.
 
 Of particular interest in this paper is how this force affects the trajectory of motion of a particle, say a spacecraft,  comet, asteroid, planetoid, etc, as it moves about the Sun sufficiently far away. If a particle is moving near the Sun, and  if it isn\rq{}t too far away, say no more than a thousand AU,  then the gravitational force of the Sun is substantially more dominant than $F_{G,S}$ on the particle. ${\bf F_{G,S}}$ can be viewed as a perturbation of the Sun\rq{}s gravitational force. 
Although this perturbation is small, it can build up over sufficient time spans. If an object is moving in a GC-centred coordinate system, then ${\bf F_G}$ is analogously considered.

 For example, if an object, say a comet,  is moving away from the Sun, at a constant velocity $v = 5$ km/s on a linear trajectory radially away from GC, then this force is acting in a direction opposite to the direction of motion.  Relative to GC it is also moving with velocity $v$ since the Sun has a zero radial velocity wrt GC, and over time, $t$, this acceleration can have an appreciable effect. It yields a resultant velocity $v - F_Gt$ relative to the GC, and also  relative to the Sun in this case.  In only 1 million years this can have a significant effect since $F_G t \approx 5.7$ km/s.  The effect of ${\bf {F_G}}$ causes a displacement or deviation of the trajectory away from the linear path. In general, the deviation of a 
 trajectory can be used to measure the effect of ${\bf{F_G}}$ (see Section \ref{sec:Test}). In this way,  {\bf $F_G$ } can be measured.
  
 This force may be important to consider when considering spacecraft missions far from the Sun.  This is discussed in general in Section \ref{sec:Test}, where it is detected from the Earth as the tidal force. The detection of the tidal force presents challenges since it is so small and the deviation of the trajectory will be difficult to detect.  A mission is proposed in Section \ref{subsec:Interstellar} that may be able to accurately measure ${\bf F_{G,S}}$. 
  
 Several examples are considered of operational spacecraft.  Two such spacecraft are the Pioneer 10, 11, launched in the early 1970s. In the early 1990s it was observed that an anomalous force was acting on the Pioneer spacecraft. Although the results of Pioneer 11 were inconclusive,  since Pioneer 11 stopped transmitting in 1995, this force on Pioneer 10 was 
accurately determined to be  $F_T = 8.74 \times 10^{-10}$m/s$^2$, with an error bar of $\pm1.33 \times 10^{ -10}$m/s$^2$, in Sun-centred coordinates. Pioneer 10 was moving approximately radially away from GC (see  \citet{Turyshev2010}(figure 2.2))   and therefore approximately perpendicular to the Sun\rq{}s circular motion about the GC.  It was estimated by Turyshev et. al.  (\citet{Turyshev2010},\citet{Turyshev2011},\citet{Turyshev2012}) that thermal properties of the spacecraft driven by its Radioisotope Thermal Generator (RTG) probably gave rise to most of this anomalous force, also pointing towards the Sun (i.e. towards the direction of GC).  $F_T$ has the same value in a GC-centred coordinate system in this case due to the Sun\rq{}s zero radial motion.  Comparing this to $F_G$  it  is seen that the magnitude of $F_G$ is approximately  $1/5$ and it is slightly outside the  error bar. Since it was close to the error bar, it wasn\rq{}t statistically  noticed,  but over time as the thermal effect decreases it may become more evident and be revealed. Unfortunately Pioneer 10 is no longer transmitting data, and cannot be used to check this. However, as discussed in Section \ref{sec:Test},  there is a problem of even detecting the tidal force  ${\bf F_{G,S}}$ to make any inferences on the value of ${\bf F_G}$ in a GC system.   The New Horizons spacecraft also measures this thermal force with a smaller error bar but  runs into the same issues on being able to detect the tidal force. 

 ${\bf  F_{G,S}}$ has implications on all objects moving sufficiently far from the Sun, as a force that should be taken into account.  This would include Kuiper belt objects, comets in the Oort cloud, Planet X. (See the discussion of Heisler and Tremaine (\citet{Heisler1986})  in Section \ref{sec:Objects} on the Oort cloud.)   These examples and others in planetary astronomy and astrophysics are discussed in Section  \ref{sec:Objects}.

The general methodology and introduction of the results of this paper are given in this section, Section \ref{sec:Intro}. In Section \ref{sec:Models} the models are derived. The application of the results to spacecraft and the problem with detection is described in  Section \ref{sec:Test}.  Applications to problems in astronomy and astrophysics, and future missions,  are in Section \ref{sec:Objects}.  Concluding remarks are in Section \ref{sec:Conclusions}.

\section{Models} \label{sec:Models} 

The two models described in Section \ref{sec:Intro} are derived. They are also shown to have nearly identical force values near the Sun and  yield valid values for the velocity of the Sun about GC.  
\medskip

\noindent
{\it Methodology}:  The goal of these models is to estimate the resultant force on a particle near the Sun mainly due to dark and baryonic matter.  The force field being generated is a central one in both models. In each case we compute the magnitude of this force at an arbitrary  point near the Sun using the force obtained from mass-density from Poisson\rq{}s equation.   This force is viewed as a perturbation of the gravitational force due to the Sun.  The two models used reflect the existence of dark matter.

 \medskip
 \noindent
   The first model is a two stepped Hernquist model, which reflects dark matter and also baryonic matter up to 300 kpc through $M_{vir}$. It ignores the relative geometries of baryonic matter in the disc,  bulge and stellar halo, and dark matter  in the dark matter halo. It functions sufficiently well at $\leq r_s$ to determine the acceleration, ${\bf F_G}$ and the Sun\rq{}s velocity,  where, as defined in the Introduction,  $r_s$ is the distance from the Sun to GC.   The second, the NFW model, is designed for dark matter, mainly in the galactic  halo, and is used together with a standard Newtonian model for the baryonic matter near $8.29$ kpc using a Newtonian point-mass potential. This is done since dark matter is mainly located hundreds of kpc  from GC, and not as prevalent at 10\rq{}s of kpc  (see \citet{Watkins2019}, \citet{Dillamore2021}). A  Newtonian model is used for the baryonic matter near $8.29$ kpc. When added to the NFW model, the value of  ${\bf F_G}$ matches the two stepped Hernquist and the Sun\rq{}s velocity is close.  These two models, although different, give close agreement. Although the two stepped Hernquist model is satisfactory for the purposes of this paper, the comparison to the NFW model with a Newtonian component offers a useful check of results. 
\medskip

\noindent
Both models yield approximately  the same force value and the Sun\rq{}s velocity about GC for each is very close, lying within acceptable values.  Other modeling could be used, but the models used here are sufficient for our results and provide consistency, covering the range in our understanding of dark matter.
\medskip

\noindent
{\it Mass Estimate of MW}
\medskip

A first step in the construction of our models is to have an estimate for $M_{vir}$, which is the mass of all matter within the galactic  halo of MW, dominated by dark matter.  Estimations of $M_{vir}$  are contained in \citet{Watkins2019}, \citet{Callingham2019}, who reference a number of other studies giving similar values. We use \citet{Watkins2019} for reference, who study the kinematics  of halo globular clusters in MW obtained from {\it Gaia}.  In \citet{Watkins2019} the estimate for the virial mass in terms of solar masses of MW is given by (\ref{eq:GalaxyMass}) 
stated in the Introduction. A density model is estimated for the various components of MW:  the nucleus, containing a super-massive  black hole at its centre, a bulge, disc and dark matter halo. All but a few percent of $M_{vir}$  is non-baryonic in nature (see \citet{Helmi2008} and  \citet{Watkins2019}).   
\medskip

\medskip
\subsection{Model A  }

The first model we consider is a Hernquist model (\citet{Hernquist1990}). This model works well for elliptical galaxies or for  bulges in spiral galaxies, e.g. the Milky Way. The Sun is on the edge of the spherical bulge.  Although this model is not designed to reflect the existence of dark matter, it is adapted for that by using $M_{vir}$ which is dominated by dark matter and by using this model in a two stepped process.  
\medskip

\noindent
The first step is for a  larger Hernquist model defined for MW in the large massive dark matter halo and the second is a smaller Hernquist model 
defined for the smaller less massive stellar halo of MW mostly of baryonic mass.
   In the first step, we consider the mass-density equation $\nu(r) = (M_{vir}/2\pi) (d/(r(r + d)^3)$ for the density $\nu$ within MW at a distance $r$ from GC. This is calculated for MW using $M_{vir}$ from (\ref{eq:GalaxyMass}) and with a scale length $d = 21.5$ kpc.   The mass contained within the distance $r$ is given by  $M(r) = M_{vir}  r^2/(r+d)^2$.  $M_{vir} = M(R_{vir})$, $ R_{vir} =300$ kpc is the virial radius of MW which is a measure of how far the dark matter halo extends. The values of $d,  R_{vir}$ are assumed from \citet{Watkins2019}.  Setting $r=r_s = 8.29$ kpc, defined in Section \ref{sec:Intro} as the Sun\rq{}s distance to GC, the mass of MW within $ \leq r_s$ is given by,

\begin{equation}
M(r_s) = .077M_{vir}.
\label{eq:HernquistMassInsideSunDistance}
\end{equation}
Thus, the mass within the sphere of radius 300 kpc for the galactic halo  of MW is used to estimate the mass within the smaller sphere of radius $r_s$.

\medskip

\noindent
The second step is to calculate the Hernquist force for $\leq r_s$.  This uses a smaller Hernquist model defined for the less massive stellar component of MW using a smaller 
scale length, $b$.   The Hernquist force for this smaller model at a point $\bf x$ within this spherical region is given by
\begin{equation}
{\bf{F}_{G,H} }({\bf{x}})=  - \frac{  GM(r_s){\bf{x} }} {r(r+ b)^2},
\label{eq:GalForce}
\end{equation}
where ${\bf{x}} = (x_1, x_2, x_3)$ is relative to $GC$,  $r = |{\bf{x}}| = (x_1^2 +x_2^2 +x_3^2)^{1/2} \leq r_s$, and where $b$ is a new scale length  for the stellar halo of MW $\leq r_s$.     $G$ is the Newtonian gravitational constant.
As in \citet{Watkins2019}, we choose $ b=1$ kpc.  This force is obtained from Poission\rq{}s equation   $\nabla^2 U(r) = 4 \pi G \nu(r)$ for the potential $U(r)$, where ${\bf F_{G,H}} = -{\bf \nabla} U$, ${\bf \nabla} = (\partial / \partial x_1, \partial / \partial x_2, \partial/ \partial x_3)$.    ${\bf{F}_{G,H}}$ is a central radial force field, per unit mass, directed towards $GC$.  The magnitude of this force is 
  
\begin{equation}
F_{G,H} (r)=   \frac{GM(r_s)} {(r+ b)^2}.
\label{eq:GalForceMag}
\end{equation}
(\ref{eq:GalForce}) can be written as,  ${\bf{F}_{G,H} }({\bf{x}}) = F_{G,H} (r) {\bf \hat x}$,  ${\bf \hat x} = {\bf x}/r $.
\medskip

\noindent
It is verified that setting $r=r_s$, then the magnitude at the Sun in a GC-centred coordinate system is
\begin{equation}
F_{G,H}(r_s) = 1.81 \times 10^{-10} m/s^2.
\label{eq:Hforcevalue}
\end{equation}
\medskip\medskip

 It is remarked that if a Hernquist force was calculated at $r=r_s$ for the larger Hernquist model, dominated by dark matter, in (\ref{eq:GalForceMag}) by replacing $b$ with $d$, then labeling this force $\tilde{F}_{G,H}$,  $\tilde{F}_{G,H}(r_s)  = [(r_s+b)/(r_s+d)]^2  F_{G,H} \approx .3  F_{G,H}.$  $\tilde{F}_{G,H}$ is substantially smaller than $F_{G,H}$ due to the different scale lengths. $b$ is an accurate scale length at the Sun\rq{}s distance from GC due to the relative high concentration of baryonic matter within the Sun\rq{}s distance to GC where $d$ is too large, and $\tilde{F}_{G,H}(r_s)$  is inaccurate.  The value of $F_{G,H}$ given by (\ref{eq:Hforcevalue}) is an accurate value.  This value is obtained in Section 2.2 for Model B. The relative low concentration of dark matter within the Sun\rq{}s distance from GC is discussed in a more precise manner at the end of Section 2.3 for Model B. 

   In summary,  the two scale lengths used in Model A are $b = 1$ kpc, $ d = 21.5$ kpc. 

\subsection{Model  B}
\label{subsec:ModelB}

Model B is the NFW model for the dark matter halo together with a point-mass Newtonian potential model for the baryonic matter for the stellar halo, labeled as the NFW+B model. 
The NFW model is a standard model for dark matter halos (\citet{Navarro1996}).  It describes dark matter halos of galaxies. This is done by assuming a mass-density function $\rho(r)$, where $r$ is the distance to GC,
$\rho(r) = \rho_0/[(X(1+X)^2]$, where  $X = r/R_s$. $R_s$ is the scale distance and $\rho_0$ is a scaled density related to the mean density within $R_s$.  A potential function ${U}(r)$ is given by Poisson\rq{}s equation,   $\nabla^2 U(r) = 4 \pi G \rho(r)$,  where,  as in Model A,  ${\bf{x}} = (x_1, x_2, x_3)$  are cartesian coordinates relative to $GC$, $r = |{\bf{x}}| = (x_1^2 +x_2^2 +x_3^2)^{1/2}$.   The acceleration or force ${\bf{F_G}} \equiv {\bf{F_{G,NFW}}}$ per unit mass,  is given by ${\bf{F_{G,NFW}}} = -{\bf \nabla} U = -\alpha h(X) r^{-3}{ \bf{x}} = F_{G, NFW}(r) {\bf \hat x}$, where 
\medskip
\begin{equation}
F_{G, NFW}(r) = \alpha h(X) r^{-2},
\label{eq:NFWforce}
\end{equation}
is the magnitude of the force, $\alpha = GM_{vir}/h(c), h(c)= \ln(1+c) - c(1+c)^{-1}$, and ${\bf \hat x} = {\bf x}/r$ is a unit vector.   We assume $R_s = d = 21.5$ kpc and $R_{vir} = 300$ kpc as in Model A  used in \citet{Watkins2019}.  $M_{vir}$ is given by (\ref{eq:GalaxyMass}).   $c = R_{vir}/R_s$ is called the concentration factor.  This yields $c=14$.
Substituting these values into (\ref{eq:NFWforce}) and setting $r = r_s$ yields the magnitude at the Sun in a GC-centred coordinate system,
\begin{equation}
F_{G,NFW}(r_s) = .84 \times 10^{-10} m/s^2.
\label{eq:NFWforcevalue}
\end{equation}
\medskip

\noindent
This value is less than (\ref{eq:Hforcevalue}) since the NFW model does not model baryonic matter. To compensate for this, a baryonic component is added on. A baryonic mass for MW is taken to be 
$ M_B = 5 \times 10^{10} M_{\odot}$  (see \citet{Licquia2013}).  Using a classical Newtonian  point-mass potential, this yields the Newtonian force model, ${\bf F_{G,B}}  = GM_B{\bf x}/r_s^3$, with magnitude,
\begin{equation}
F_{G,B}(r_s) = 1.01 \times 10^{-10} m/s^2.
\end{equation}
Adding $ F_{G,B}$  to $ F_{G,NFW}(r_s)$  yields,
\begin{equation}
 F_{G, NFW+B} = 1.85 \times 10^{-10} m/s^2.
 \label{eq:ForceNFWB}
\end{equation}
These two force magnitudes can be added since  {$\bf F_{G,NFW}$} and {$\bf F_{G,B}$} are both radial central force fields.   
\medskip

 The value of $ F_{G, NFW+B}$ in Equation \ref{eq:ForceNFWB} is in close agreement with the value of $F_{G,H}$  in (\ref{eq:Hforcevalue}). 
Also, both ${\bf F_{G,NFW+B}}$ and ${\bf F_{G,H}}$ are central radial force fields, per unit mass, directed towards $GC$, increasing in magnitude as $r$ decreases.   
\medskip

It is concluded that Model A is sufficient for the purposes of this paper, which is checked against Model B. This is also verified in the next section on the computation of the Sun\rq{}s velcoity.  
Thus, we make the following assumption for the remainder of this paper,
\medskip

\noindent
{\it Value of ${\bf F_G}$}
\medskip\medskip

${\bf F_G}= {\bf F_{G,H}}$ with magnitude at $r = r_s$ given by (\ref{eq:FGValue}). 
\medskip\medskip

 It is remarked that other models could be considered (see \citet{Bovy2014})  which is out of scope of this paper and for future study.  

\medskip
\subsection{Computing the Sun\rq{}s Velocity}

These two models are checked to see if they can accurately compute the velocity of the Sun about GC. 

The circular velocity of the Sun about GC  is defined to be,
\begin{equation}
V_{C,s} = \sqrt{\frac{GM( r_s)}{r_s}}.
\label{eq:CircularVelocity}
\end{equation}
This can be computed for Model  A, where $M(r_s)$ is given by (\ref{eq:HernquistMassInsideSunDistance}). This yields,
$V_{C,s} = \sqrt{.077}\sqrt{GM_{vir}/ r_s}$,
\begin{equation}
V_{C.s} = 248 \:  km/s.
\label{eq:HCircularVelocity}
\end{equation}
This is an acceptable value lying within the admissible region of the estimated circular velocity of the Sun, 220-250 km/s, the same range used in  \citet{Watkins2019}. 
\medskip

 This model  incorporates both baryonic and dark matter through $M_{vir}$. It doesn\rq{}t make a distinction of the geometries of these two types of matter.  This yields an accurate value for both the Sun\rq{}s velocity and also the acceleration due to both these mass contributions. 

To compute the Sun\rq{}s velocity for Model B, we first need to compute $M(r_s)$ for the NFW model for dark matter predominately in the halo.
 $M(r_s)$ is computed using the mass-density,
\begin{equation}
M(r_s) = 4\pi \int^{r_s}_0 \rho(r)r^2dr .
\end{equation}
 This can be expressed as, 
\begin{equation}
M(r) = M_{vir}\frac{h(c\tilde{X})}{h(c)},
\end{equation}          
where $\tilde{X} = r/R_{vir}$ and then setting $r=r_s$.  This yields,
           \begin{equation}
 M( r_s) = .029 M_{vir}.
\label{eq:MassInsideSunDistance}
\end{equation}
Computing (\ref{eq:CircularVelocity}) in this case, we obtain,  $V_{C,s} = 152$ km/s.  This is labeled $V_{C,s}^{NFW}.$
This value lies significantly below the interval 220-250 km/s.  This is because the model is designed to model dark matter whose concentration at the Sun\rq{}s distance from GC is relatively small.
\medskip

\noindent
 Since baryonic matter is more dominant at the Sun\rq{}s distance from GC,  its mass value for MW is added to the NFW mass value $M(r_s)$ given by (\ref{eq:MassInsideSunDistance}),

\begin{equation}
M(r_s) = M_{NFW+B}(r_s)  = M_B + .029 M_{vir}.
\label{eq:TotalMass} 
\end{equation} 

Substituting $M(r_s)$ into  (\ref{eq:CircularVelocity})  yields the modified circular velocity equation at $r=r_s$ for the Sun for the  NFW+ B model, $V_{C,s}= \sqrt{ [{V_{C,s}^B}]^2 + [{V_{C,s}^{NFW}}]^2}= \sqrt{.0615}\sqrt{G M_{vir}/r_s}$, where $V_{C,s}^B = \sqrt{GM_B/r_s}$.  The Sun\rq{}s velocity is calculated to be,
\begin{equation}
V_{C,s}=222 \: km/s,
\end{equation}
which also lies within the acceptable range.
\medskip

 \noindent
 The final results for Models A, B are summarized in Table \ref{tab:Table1}.
 \medskip\medskip

\noindent
{\it  Dark and Baryonic Matter from Model B}: 
\medskip

  \noindent
It is noted that for Model B,  the value of $F_{G,NFW + B}(r_s)$ is due to about $45\%$ from dark matter and $55\%$  from baryonic matter at the Sun\rq{}s distance from GC as seen from $F_{G,NFW}, F_{G,B}$. This is consistent with \citet{Eilers2019} (see Figure \ref{fig:fig1}).  In the entire MW, it is seen that $M_B$ is about $3\%$ of $M_{vir}$.   
\medskip

\noindent
Thus, the relative force contributions per unit mass from baryonic and dark matter at the Sun\rq{}s distance from GC, from Model B, respectively, are about
$F_{baryonic} =   1.0  \times 10^{-10}$ m/s$^2$, $F_{dark} = .8 \times 10^{-10}$ m/s$^2$.

\begin{table}
\caption{Acceleration and Sun\rq{}s velocity from models  at $r_s$, and where $R_{vir} = 300$ kpc}.
\centering
\begin{tabular}{||ccccc||}
\hline
Model  & Symbol &  Equ.  &   Mag.(m/$s^2$) & $V_{C,s}$(km/s)  \\ [1ex]
\hline
 A  &  $F_{G,H}$  &  \ref{eq:Hforcevalue} & $1.81 \times 10^{-10}$ & 248\\
\hline
  B   & $F_{G, NFW + B}$ & \ref{eq:ForceNFWB}  &   1.85  $ \times 10^{-10} $  &  222  \\
\hline
\end{tabular}
\label{tab:Table1}
\end{table}
\medskip\medskip

\noindent
\subsection{Estimation of  the Distance from the Sun where  ${\bf F_{G,S}}$ becomes Dominant to ${\bf F_s}$}
\medskip

\begin{figure}
\centering
           \includegraphics[width=\columnwidth]{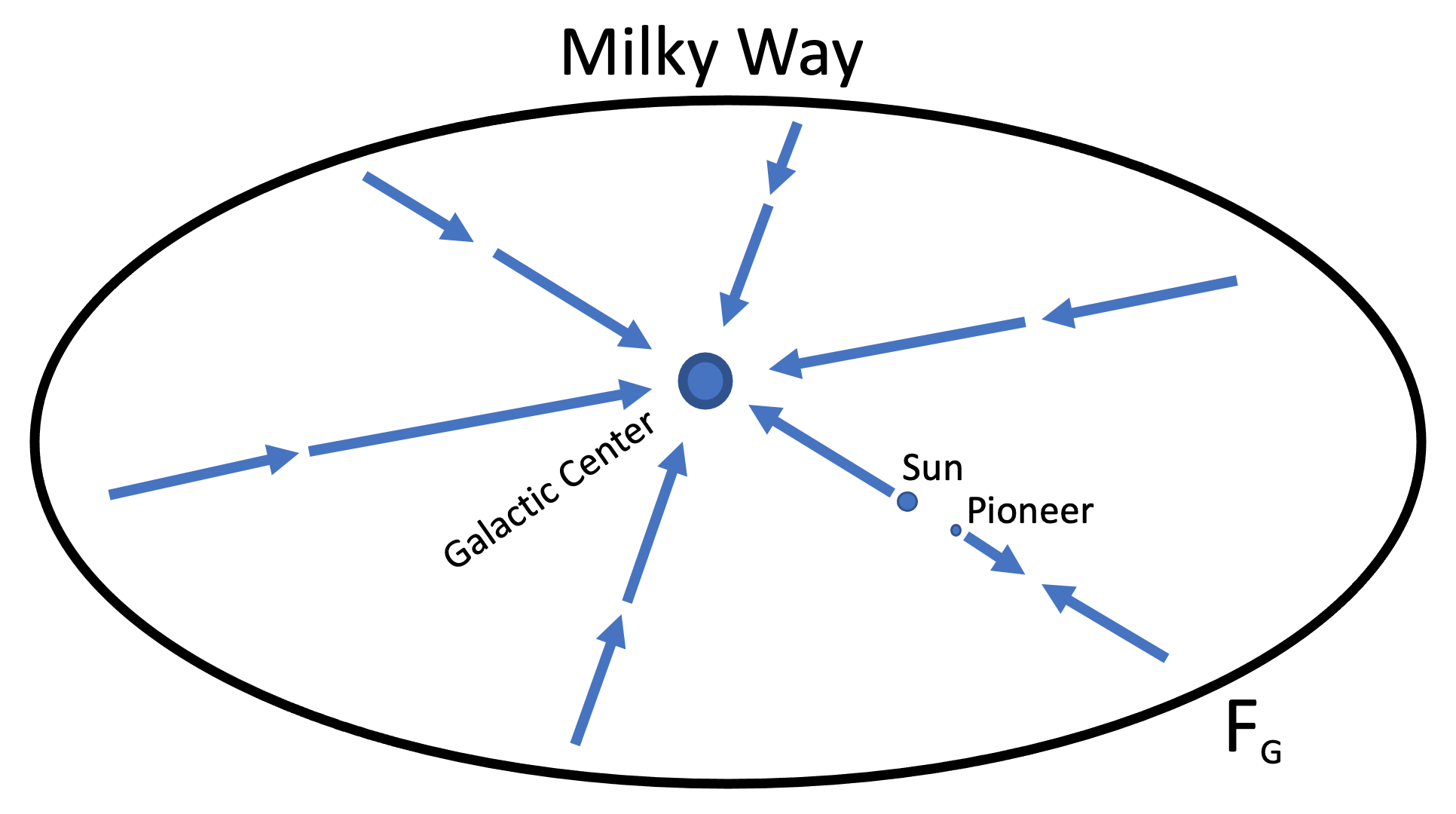}
	\caption{Motion of Pioneer radially away from $GC$ and direction of  ${\bf{F_{G}}}$ towards $GC$.}
	\label{fig:fig2}
\end{figure}

Section 2  is concluded by roughly estimating a distance from the Sun, beyond which the dominant force is $F_G$ in Sun-centred coordinates, labeled  ${\bf{F_{G,S}}}$,   relative to ${\bf F_s}$, 
the gravitational force due to the Sun in Sun-centred coordinates. This is done for an object $P$ moving away from the Sun, $S$, on a hyperbolic trajectory. It is assumed that $P$ is moving within  $1$ LY from $S$ within the stellar halo. 
For simplicity of argument, we assume that the trajectory, which is approximately rectilinear far from $S$, moves radially away from $S$,  $180$ degrees from $GC$. This is approximately the case for Pioneer 10.  
 Thus, $P$  moves radially outward from GC in the $GC$, $S$ - plane. In a GC-centred system ${\bf F_G}$ points radially inward to GC  (see Figure \ref{fig:fig2}).  In a Sun-centred system ${\bf{F_{G,S}}}$ points in a direction given by (\ref{eq:TidalEqu3}), 
 which is a vector of small magnitude at $P$, pointing radially away from $P$  towards the direction of the Sun. Its magnitude is discussed in the next section.

   The motion of $P$ is modeled as a rotating two-centre problem, since $S$ is rotating slowly within MW relative to $GC$.  One centre is $GC$ with mass $M_{vir}$ and the other gravitational centre point is $S$ with mass $M_{\odot}$. 
   The acceleration field of MW near $S$ is given by ${\bf{F_G}}$. Thus, we consider $P$ moving within the acceleration field generated by these two rotating centre points.  Since the motion is approximately radially outward from $GC$, ${\bf{F_G}}$  
   acting on $P$ along this direction is given as a one-dimensional system with $y$ being the coordinate along the radial direction from $S$. Thus, $r = r_s + y$,  where $r_s = 8.29$ kpc. 
 \medskip
 
  \noindent 
   We will put the forces in terms of $y$ to have a Sun-centred coordinate system.     This yields a differential equation for its effect on $P$.  As described in the next section, a tidal force, $F_{G,S}$,  
   for $F_G$ needs to be considered since the acceleration field for the galactic force acts on both $P$ and the Sun. Thus, 
\begin{equation}
F_{G,S} ={ \ddot{r}}-{\ddot{r}_s} =   {GM(r_s) }[(y+ r_s+1)^{-2}  - ( r_s +1 )^{-2}],
\label{eq:DEGalaxy}
\end{equation}
where $M(r_s)$ is given by (\ref{eq:HernquistMassInsideSunDistance}).  It is noted that the Hernquist model in the second step requires $r \leq r_s$. However in this case we are slighly beyond this distance at $r_s + y$. But since $y \approx 0$ relative to the GC, (\ref{eq:DEGalaxy}) is valid.
\medskip

\noindent
On the other hand, also acting on $P$ relative to the Sun is the Newtonian point-mass gravitational force, $F_s$, per unit mass,
\begin{equation}
 F_s =   -\frac{GM_{\odot} }{y^2} 
\label{eq:DESun}
\end{equation}
A value of $y > 0 $ is sought where $F_{G,S}  = F_s$. To make this equality,  (\ref{eq:DEGalaxy})  is reduced by a binomial expansion; 
\begin{equation}
(y+(r_s+1))^{-2}  = (r_s + 1)^{-2} [1 + (y/(r_s+1)) ^{-2}] 
\end{equation}
\begin{equation}
  =  (r_s + 1)^{-2}[1-2(y/(r_s+1) + \mathcal{O}((y/(r_s+1)^2)],
  \end{equation}
where $(y/(r_s+1)$ is much smaller than $1$. It is noted that in the units of kpc, that $r_s$ is measured, $y$ is small and much less than $1$.
Substituting this into  (\ref{eq:DEGalaxy}) and equating to $F_s$, yields,
\begin{equation}
y^3 = [(r_s +1)^3/(.2 \times 10^{12}) )]   + \mathcal{O}(10^{-12} y^4).
\end{equation}
This yields,
\medskip

$y \approx 31,000$ AU.
\medskip\medskip

\noindent
It is important to note that this number is much larger than where this force would actually be felt since the calculation is only saying where it dominates the Sun\rq{}s gravity. Since ${\bf F_G}$ gives an acceleration field that exists throughout our solar system,
 it should be felt as soon as a spacecraft leaves the Earth.  However, as described in the next section, it is a question of detecting this force. If it could be measured in GC-centred coordinates, then its magnitude could be detected more easily. 
 But as described in the next section, $F_{G,S}$ is small and would not be easy to detect. 
\medskip\medskip

The previous analysis estimates where the magnitudes of the forces ${\bf F_{S,G}, F_s}$ are equal 
for $P$ in the given model. 
However, from a dynamics perspective  it would be interesting to analyse the motion of $P$ where all the forces on $P$ are all approximately balanced.  To do this, the centripetal force is included. This force,  ${\bf C_s}$, is defined in the next section relative to the Sun, ${\bf C_s} = \omega_s^2{\bf x_s}$,  where $\omega_s$ is the rotational velocity of $S$ about GC. 

There is a method that could analyse the dynamics of a particle $P$  moving about the Sun in regions where the forces ${\bf F_{G,S}}, {\bf F_s}, {\bf C_s}$  acting on $P$ approximately balance and could give rise to a sensitive motion between capture and escape about the Sun. Such regions, called weak stability boundaries,  have been studied in detail
for motions about the Earth\rq{}s Moon and used in applications to find trajectories for operational spacecraft to be captured about the Moon with minimal energy.\footnote{ Lunar missions using low energy capture trajectories  include {\it Hiten} of ISAS (1991) (\citet{Belbruno2004}), {\it  SMART-1} of ESA (2003) (\citet{Racca2002}), {\it GRAIL} of NASA (2011) (\citet{Roncoli2012}).}  These regions are called weak stability boundaries, and they have a complex fractal structure that is far from spherical  (\citet{Belbruno2010}).  They are determined with a numerical algorithm by estimating transitions between capture and escape trajectories about a body, perturbed by the gravitational force of a larger body. A boundary is graphically shown about Jupiter in \citet{Topputo2009}.  The analysis of such a region about the Sun is beyond the scope of this paper and a topic of future study.

\medskip\medskip\medskip

\section{Detection of the Force ${\bf F_G}$  Due to Dark and Baryonic Matter by a Spacecraft} \label{sec:Test}

The value of $F_G$  is small.  Can this be detected?  
\medskip\medskip

 In practice, to detect the acceleration $ F_G$  one has to measure the effect of $F_G$ on an object.  This can be done by measuring the displacement of the object. 
If the object is moving, then the effect of $F_G$ will be to cause a deflection of its trajectory from  Keplerian motion.  Because $F_G$ is small, the deflection is small.  
In fact, in a Sun-centred coordinate system, $F_G$ is determined as a tidal force, $F_{G,S}$,  which is even much smaller, as described in this section.  Relative to the Sun, the deflection will be even smaller. However, it can be measured. 
A mission is described at the end of Section \ref{subsec:Interstellar} that may be able to do this.
\medskip

We consider a coordinate system centred at the Sun,$S$, with coordinates ${\bf y}=(y_1, y_2, y_3)$.  For a particle, $P$, say a spacecraft, moving with respect to $S$, far from any planetary bodies, but relatively near to the Sun within $1$ LY, there are three forces acting on $P$.  They are:
\medskip
 
  \noindent
    1.) {\it The gravitational force due to the Sun, $\bf{F_s}$ }  
    \medskip 
    
      This force per unit mass is given by 
$$ \bf{F_s}  = -  GM_{\odot}{\bf y}/|{\bf y}|^3.$$  
in a Sun-centred coordinate system ${\bf y}$.   It acts in the direction of the Sun.
\medskip

\noindent
Letting ${\bf x}$  be the vector coordinates for $P$ in a GC-centred system, and ${\bf x_s}$  the vector coordinates of the Sun relative to GC, then
\begin{equation}
{\bf x}  = {\bf x_s} \bf{+} {\bf y}.
\label{eq:Coordinates}
\end{equation}

\noindent
2.)  {\it  The centripetal force  ${\bf C_s}$ due to the Sun\rq{}s circular motion about GC}   
\medskip

It is directed towards GC. It is given by 
$$ { \bf{C_s}} = -\omega_s^2{\bf x_s},$$ where $\omega_s$ is the rotational velocity of $S$ about GC.  It is calculated that
the magnitude of this force is given by,
$C_s =   2.34 \times 10^{-10}$m/s$^2$.   This force  is larger than $F_G(r_s)$.    In a Sun-centred coordinate system this force value acts on $P$ , and has value 
$ + {\bf \omega_s^2{\bf x_s}}$, as follows from (\ref{eq:Coordinates}).
\medskip\medskip

\noindent
3.) {\it  ${\bf F_{G}}$ } 
\medskip
 
   In a Sun-centred coordinate system, ${\bf F_G}$ at a point ${\bf x}$ near the location of the Sun, ${\bf x_s}$,  is given by the tidal force
\begin{equation}
{\bf F_{G,S}}({\bf y}) = {\bf F_G}({\bf x}) {\bf -} {\bf F_G}({\bf x_s}),
\label{eq:TidalEqu}
\end{equation}
where ${\bf x_s}$ , ${\bf x}$ are in a GC-centred coordinate system, as follows from (\ref{eq:Coordinates}).  The direction of ${\bf F_{G,S}}({\bf y})$ depends on the location of $P$.
 \medskip
 
 \noindent
Since $\bf F_G$ is a central force field with magnitude $F_G$, then Model A or B implies (\ref{eq:TidalEqu}) can be written as, 
\begin{equation}
{\bf F_{G,S}} ({\bf y}) =F_G(r){\bf \hat x} - F_G(r_s){\bf \hat x_s}. 
 \label{eq:TidalEqu2}
 \end{equation}
   Since ${\bf \hat x} $ ,  ${\bf \hat x_s}$ have nearly the same direction relative to GC, and $r, r_s$ are very close in value relative to GC,  then $|{\bf F_{G,S}} ({\bf y})| $ is small.  
   \medskip

As an example, we consider the case of Pioneer 10, labeled P10, launched in 1972.  The last reception of Pioneer 10\rq{}s signal was on January 22, 2003 (\citet{Mewhinney2003}), at a distance of 82 AU from the Sun. As discussed in Section \ref{sec:Intro}, an anomalous  force value as $F_T=
 8.74 \times 10^{-10}$ m/s$^2$  was measured on Pioneer 10, in a Sun-centred system.   This thermal force value is obtained by factoring out the other forces on the spacecraft:  ${\bf{F_s}},{ \bf{C_s}}$.  This value is large enough to cause the trajectory to deviate significantly from its nominal path by approximately 380,000 km in 30 years.    Turyschev et. al. (\citet{Turyshev2012})  determined that this force was mainly due to thermal pressure from degradation of the Plutonium in the RTG. The error bar for the determination of $F_T$ is $\Delta_{E}( P10) = \pm 1.33 \times 10^{-10}$ m/s$^2$  (see Table \ref{tab:Table2}). 
 \medskip
 
 \noindent
As described in the Introduction, P10 was moving approximately radially away from GC and in a GC-centred coordinate system,  where  $F_T$ has approximately the same value in GC-centred coordinates and can be directly compared to $ F_G(r_s)$. (The value of $F_G(r)$  where $r  =  r_s + 82$ AU, and $82$ AU is converted to kpc,  is only slightly smaller than $ F_G(r_s)$ and we compare to $ F_G(r_s)$ for convenience.) $F_G(r_s)$ is about 4.9 times less than $F_T$.  
\medskip\medskip\medskip

\noindent
{\it \bf A Problem of Detection, Relative Coordinate Systems}
\medskip

\noindent
 ${\bf F_G}$ has a magnitude of $1.8 \times 10^{-10}$ m/$s^2$  in a GC-centred coordinate system at the location of the Sun.  If a location is within say a few hundred AU from the Sun, for a spacecraft, this force will be slightly different, but nearly the same.  Its difference in magnitude will be much smaller than $F_G$ itself.   But it is necessary to consider this difference, or tidal force  ${\bf F_{G,S}}$,  in a Sun-centred system.  This is because both the Sun and the spacecraft are within the force field(or acceleration field).  Thus, when actually detecting this force on a spacecraft, or any other object relatively near to the Sun, it must be computed.   If an object is far from the Sun, say many light years, then  ${\bf F_{G,S}}$ may not be so small in magnitude and the deviation of the trajectory more easily detected.

 This is an important issue, since no matter what is observed it must be done relative to our solar system,  Sun-centred,  for it to be estimated.  In the case of rectilinear motion radially away from the Sun, opposite to the direction of the GC, as approximately for P10,
 ${\bf \hat x }    \approx    {\bf \hat x_s}$.  Thus, 
\begin{equation}
{\bf F_{G,S}} ({\bf y})  \approx [F_G(r) - F_G(r_s)]{\bf \hat x_s}. 
\label{eq:TidalEqu3}
\end{equation}
Thus,  $|{\bf F_{G,S}} ({\bf y})| =|F_G(r) - F_G(r_s)|$. Using Model A, this can be approximated as, 
\begin{equation}
|{\bf F_{G,S}} ({\bf y})|   \approx  F_G(r_s)  \frac{2\delta}{r_s + 1},
\label{eq:TidalEqu4}
\end{equation}
where $\delta  = 82$ AU in kpc  $\approx 3.9 \times10^{-7} $ kpc.  This gives,  $$|{\bf F_{G,S}} ({\bf y})|   \approx .8 \times 10^{-7}   F_G(r_s) . $$
\medskip\medskip

\noindent
Thus, in a Sun-centred system, under the assumption of rectilinear motion away from the Sun, opposite to  the direction to the GC, 
\medskip 

\noindent
  {\it The magnitude of the galactic force at {\bf y}, ${\bf F_{G,S}} ({\bf y})$,  where P10 is located, is only $.8 \times 10^{-7}$ of its value at the Sun relative to a GC-centred system.}
  \medskip
  
\noindent  
This estimate can be made for any general direction of motion from the Sun. This implies that detecting this force from the Earth may be difficult since it is roughly 10 million times less than $F_T$, and $F_T$ was already at the limits of detection.   However, it may be done by carefully measuring the deviation of the trajectory over sufficiently long periods of time. In the case of P10, based on the deviation of the trajectory of approximately 380,000 km in 30 years due to $F_T$, this implies that  relative to the Sun,  ${\bf F_{G,S}}$ would roughly cause a deviation of only 1.6 meters, assuming a direct scaling.  If such a tiny deviation were measured, this could infer the magnitude $ |{\bf F_{G,S}} ({\bf y})|$ in Sun-centred coordinates which would yield $F_G$  in GC-centred coordinate system, and could be compared to $F_T$. If $F_T$ were less than this force, then maybe  $F_{G,S}$  could be detected.  It is noted that P10 is no longer communicating so that this strategy cannot be carried out.
\medskip

\noindent
Even though a deviation of approximately 1.6 meters  in 30 years is small, for objects such as comets, relative to the Sun, the effects would be more significant for sufficiently long times. As a comet gets sufficiently far from the Sun, ${\bf F_{G,S}}$ may not be so small, and therefore the trajectory may deviate more and the force more easily detected.  
\medskip\medskip

It is noted that when transforming the velocities of an object $P$ moving away from the Sun, say in a rectilinear fashion radially from the direction of GC, that the velocity of $P$ relative to the Sun, due to ${\bf F_{G,S}}$, is the same as relative to GC due to ${\bf{F_G}}$.  It isn\rq{}t smaller as in the case of the  force magnitude of ${\bf F_{G,S}}$ when compared to the magnitude of ${\bf F_G}$. This is because the velocity change of the Sun due to ${\bf{F_G}}$ is zero, since the Sun is moving on an approximate circular orbit about GC.   Thus, (\ref{eq:Coordinates}) yields,  ${\bf \dot{x}} = {\bf \dot{y}}$. 
\medskip\medskip

It is remarked that since Pioneer 10 is  moving away from the Sun in a direction approximately opposite to the direction to GC, then  from the perspective of the Sun, ${\bf F_G}$ that is acting on Pioneer 10 is pointing towards the Sun. The Pioneer anomaly was also pointing towards the Sun, but for different reasons   (see  Figure \ref{fig:fig2}).   It is also seen that in this case ${\bf F_{G,S}}$ approximately points in a direction radially towards the Sun.
\medskip

\medskip
\medskip

There is another spacecraft which may give an opportunity to measure $F_G$. This is the New Horizons spacecraft which we now consider. It has the same issues for the detection of the tidal force.
\medskip

The New Horizons spacecraft was launched on January 16, 2006.  The thermal force was measured to be  $13.2 \times 10^{-10}$m/s$^2$ in 2008 when it was at 8AU, which is slowly decreasing with the RTG (\citet{Rogers2014}, \citet{Guerra2017}). The error bar in this mission is  lower than Pioneer 10. It is $\Delta_{NH} = \pm .6 \times 10^{-10}$ m/s$^2$  (see Table \ref{tab:Table2}).\\

\begin{table}
\caption{Measured $F_T$ (the observed thermal force)  and the error bars.}  
\centering
\begin{tabular}{||cccc||}
\hline
Spacecraft  & Distance (AU) & $\Delta_E$ (m/s$^2$) & $F_T$ (m/s$^2$)  \\ [1ex]
\hline
P10 (2003)  & 82 & $1.33 \times 10^{-10}$ & $8.74 \times 10^{-10}$\\
\hline
NH (2008)  & 8 &  $.6 \times 10^{-10} $ &$13.2 \times 10^{-10}$ \\ [.1ex]
\hline
\end{tabular}
\label{tab:Table2}

\end{table}

\noindent
The New Horizons spacecraft was well designed to reach Pluto and into the Kuiper belt (30-60 AU), but not designed to be fully functional at twice that distance.  The spacecraft is about the size of a grand piano, in a triangular shape with the $+x$-axis containing one RTG.  The RTG is decaying exponentially with the power decrease currently about 3.2 W/y starting with a power at the beginning of the mission of 246 W. Flexibility in New Horizons operational modes provides the capability to downlink data using its 12W RF transmitter when the RTG power is only 105 W, but the spacecraft propellant may freeze before the RTG reaches that power level. Therefore, to successfully contact the spacecraft at its furthest distance, it is critical that a health assessment be made each year to determine when last contact could occur and adjust the spin axis to maximize opportunities for future contacts in the event that subsequent adjustments become impossible. Ranging and delta DOR measurements, which are used to determine the position of the spacecraft, may continue to the end of the mission, if required, but DSN contacts will be necessary in any case for the above health assessment to be made. Assuming no critical anomalies occur, the last contact date could be as late as 2052. With New Horizons currently travelling at 13.86 km/s with respect to the Sun this puts the expected maximum possible distance to be about 140 AU  (see Table \ref{tab:Table3}).
\medskip

\noindent
During active periods, such as the Pluto and Arrokoth flybys, New Horizons\rq{} spin axis ($+y$-axis through the centre of the radio dish) was set to follow the Earth. During hibernation periods, the spacecraft spin axis and radio antenna axis were set to the position where the Earth will be at the time of spacecraft wake-up.  This leaves the RTG, and therefore, the thermal force at right angles to the Sun-spacecraft line. At this time, no pointing strategy has been developed near the end of the mission.

\begin{table}
\caption{Maximum Distance of P10, NH. }
\centering
\begin{tabular}{||cccccc||}
\hline
  & $ P/P_0$    &  BOM  & Decay & Last & Distance \\             
 &  10 years  & watts  & watts/yr  & Contact  & AU \\
\hline 
P10  & 20 $\%$ &160 & 3.2 $\%$ & 2003 & 82 \\
\hline
NH  &  $\approx$10 $\%$  &  246  & 2.46-3.2 $\%$ & 2052 est&  139.7  \\ 
\hline
\end{tabular}
\label{tab:Table3}

\end{table}

Is it possible to detect the existence of $F_G$ from the New Horizons spacecraft?  The following analysis shows that it  may not  be possible:   It is estimated in \citet{Guerra2017} (see Fig 6) that the thermal acceleration, $F_T$, on the NH spacecraft due to the RTG has the approximate minimum value of  $3 \times 10^{-10}$m/s$^2$ 20 years after launch in 2026. This value is 2 standard deviations from the mean. This is about  $2.4 \times 10^{-10}$m/s$^2$ above the error bar,  well above the value of $F_G$ that could be felt by NH, in GC-centred coordinates (assuming deviations in the trajectory could be measured in Sun-centred coordinates to deduce $|{\bf F_{G,S}} ({\bf y})|$, then transformed to GC coordinates).  The thermal acceleration magnitude, $F_T$  decays exponentially and it can be estimated at the theoretical maximum distance from the Sun in 2052, 26 years later.  The exponential factor of decay of this force is given by 
\begin{equation}
f(t) = e^{-t \log(2)/39.1},
\end{equation}
where $t$ is time in years (\citet{Guerra2017}).  This yields $f(26) = .8186$.  Thus, the minimum value of $F_T$ is approximately $2.5 \times 10^{-10}$m/s$^2$ or about $ 1.9 \times 10^{-10}$m/s$^2$ above the error bar. This is still slightly above the possible maximal value of $F_G$. Therefore, $F_G$ is slightly out of range to be felt.
\medskip

\noindent
Thus, with a 2 standard deviation minimum from the nominal thermal acceleration, $F_G$ is slightly out of range to be noticed.   As with P10, for the time ranges considered,  $F_{G, S}$  could not be detected since it would be so small. Thus, no comparison\rq{}s to $F_G$ can be made.
\medskip

\noindent
{\it Detection of Dark Matter}
\medskip

\noindent
 It is noted that once $F_{G,S}$ is detected by a deviation of a trajectory of a spacecraft, there is the problem of discerning the relative contribution of the deviation from dark and baryonic matter.  This is the case since most of the dark matter is within the large dark matter halo, most of which is beyond the stellar halo where the Sun is located.  It was seen at the end of Section 2.4 for Model B, that the approximate relative percentages of dark and baryonic matter near the Sun contributing to $F_G$ are $45 \%$ and $55 \%$ respectively. This would be different much further away from GC at a few hundred kpc where dark matter is all but a few percent of the total matter.  This implies that for detecting dark matter by measuring the 
deviation of a trajectory of a spacecraft near the Sun, the relative makeup of dark to baryonic matter needs to be more precisely understood.  This is a topic for future study.
\medskip\medskip

Is it possible to design a mission to detect the tidal force $F_{G, S}$  ?   This is discussed in the next section.

\section{Discussion: Implications} 
\label{sec:Objects}

The theory presented in this paper has far reaching effects for not only distant Solar System objects but also will play an important role in planetary astronomy and astrophysics, briefly discussed in the following subsection. A new mission is also discussed that may be able to detect ${\bf F_{G,S}}$.

\subsection{Motion of Objects in the Distant Solar System}
\medskip

\noindent
{\it Planet 9}
\medskip\medskip

Over the last decade scientific evidence with simulations have indicated that a large Neptune sized planet was created, during the early formation of the solar system, and then during the period of the giant planet migration, about 4 Ga ago, was scattered into the distant solar system. This planet is known by the name of Planet 9 or Planet X with a mass of 5-10 Earth masses. Simulations indicate that Planet 9 could have been scattered initially into a highly eccentric orbit and over time, by an unknown process, was circularized to about 400-800 AU 
(\citet{Batygin2019}). Today the clues as to the current location of Planet 9 are found in the most distant Kuiper belt objects (KBO), which have been perturbed into highly elliptical orbits, within approximately the same plane, and have an unexpected clustering in their arguments of perihelion that can only reasonably be explained by a dynamical interaction with the proposed Planet 9 (\citet{Batygin2016}). Analysis of the scattered KBOs led to a prediction as to the location in the sky where Planet 9 maybe found to be near the galactic plane in the direction of Orion.

This proposed Planet 9 is orbiting the Sun at such a distance that ${\bf F_{G,S}}$ may have a substantial effect on its motion over time. Several aspects should now be reconsidered. Assuming the nominal magnitude, $F_G$ = 1.8 x 10$^{-10}$ m/$s^2$  (GC  centred) and the length of time it has been in the distant solar system of $\sim$ 4 Ga, then it could have moved into an orbit that is substantially different than expected or could have even escaped the solar system all together. The analysis of this problem requires a lengthy analysis that is beyond the scope of this paper.   
\medskip\medskip

\noindent
{\it \lq{}Oumuamua}
\medskip\medskip

Discovered on October 19, 2017, \lq{}Oumuamua was the first interstellar asteroid sized object found transiting our solar system at speeds reaching 87.3 km/s during its closest approach to the Sun. This object is cigar shaped and is approximately 400 m long but only 40 m wide and spins on its axis every 7.3 hours (\citet{Meech2017}). Unlike the typical solar system asteroid, \lq{}Oumuamua is very dense, believed to be mostly composed of rock and possibly metals and that its surface was reddened due to the effects of irradiation from cosmic rays over hundreds of millions of years. \lq{}Oumuamua demonstrated that other star or solar systems may be regularly ejecting small bodies and that there should be many more of them drifting among the stars. Current analysis indicates that there may be at least one interstellar object within 1 AU of our Sun at any one time (\citet{Meech2017}). Current ground- and space-based telescope surveys are now on the lookout for more of these interstellar objects. Although the object will end up with about the same speed with which it entered the solar system, only its direction will have changed.

\lq{}Oumuamua entered the solar system from the general direction of the constellation Lyra with an observed velocity of 26.3 km/s presumably ejected from a nearby stellar system. Bailer-Jones et. al.  (\citet{Bailer-Jones2018}) looked for a plausible stellar system origin for \lq{}Oumuamua using Gaia data of precise stellar locations of 20 stars that \lq{}Oumuamua is expected to have  passed within 1 pc every Myr.  Adding to the problem is trying to determine how \lq{}Oumuamua would obtain the observed high velocity upon entering the solar system. Based on their analysis, the authors state that it would be unlikely that our current search would find \lq{}Oumuamua\rq{}s home star system, in addition to the fact that none of the top four candidate systems have known exoplanets. 

Using the analysis presented in this paper, an estimation of the maximum acceleration can be calculated. Under the best conditions where \lq{}Oumuamua heads for the {\it GC}, the accumulated $\Delta$V due to ${\bf F_G}$ is linear ($\Delta$V = 1.8 x 10$^{-13}$ km/$s^2$  x  T, T = time duration  in seconds) (solar system centred). If it started at rest with respect to the {\it GC}, with no stellar encounters, \lq{}Oumuamua would take only 4.64 million years to reach the accumulated velocity of 26.3  km/s. If it moved in the opposite direction to the {\it GC}, after a stellar encounter, then it would slow down.  If it moved transverse to ${\bf F_G}$ then the trajectory would deviate toward the {\it GC}. A new analysis utilizing the derived galactic force, as presented here, may lead to finding a likely origin. 

Soon after \lq{}Oumuamua\rq{}s perihelion passage, a detailed analysis of its trajectory found a non-solar acceleration directed radially away from the Sun. Since \lq{}Oumuamua was still well situated within the domain of the planets and under the Sun\rq{}s gravitational influence, this acceleration, which is significant, cannot be due to ${\bf F_G}$ but is most likely due to outgassing, a feature common to comets (\citet{Micheli2018}). 
\medskip\medskip

\noindent
{\it Oort Cloud}
\medskip\medskip

The most distant feature of our solar system is the Oort Cloud. It is believed that a cloud of small icy comets, ejected during the early formation process of the inner solar system, resides at an enormous distance from the Sun ranging from 20,000 AU to perhaps 150,000 AU in a giant spherical shell (\citet{AHearn2006}). The key evidence for the existence of such a cloud comes from the analysis of their orbits during perihelion passages ($<$5 AU) when they become visible. On the average, there is about one Oort Cloud comet that enters the inner solar system per year with a highly eccentric orbit from virtually any heliospheric latitude. 

The outer limit of the cloud has been difficult to determine due to gravitational interactions by stars that pass close to the Sun that are expected to perturb these comets inward. However, the analysis presented here clearly shows ${\bf F_{G,S}}$ must be taken into account at great distances from the Sun and that the Oort Cloud may not reside in the expected spherical shell. A comet in the Oort cloud with a distance of 100,000 AU will feel the acceleration of ${\bf F_{G,S}}$   much more than the gravitational acceleration to the Sun. Thus, a comet will deviate from its orbit about the Sun and be acted on in a dominant way by the galactic acceleration. The path of the comet will likely deviate to such an extent that the comet may move away from the solar system or toward the Sun depending on its position. 

A new analysis using observed Oort Cloud comets that takes into account ${\bf F_{G,S}}$  is warranted and would provide a truer picture of the structure of these most distant solar system objects. 
In fact, an analysis of comet motions in the Oort cloud with a general potential model for MW is contained in the paper by Heisler and Tremaine (\citet{Heisler1986}) and shows significant deviations in their motions due to the galactic tidal force perturbations. The potential model used in  \citet{Heisler1986} is different than the model used in this paper.
\medskip\medskip

\subsection{Motion of Objects Outside the Solar System}
\label{subsec:Interstellar}
\medskip

The ultimate source of solid material in building solar system objects, from rocky planets to asteroids, begins with micron or sub-micron sized cosmic dust.  The accretion of dust particles leading to progressively larger and larger objects is not completely understood. Cosmic dust is created in a variety of ways, from condensing interstellar clouds, stellar explosions, to forming in the cooler outer layers of large red giant stars and then carried into the interstellar medium by the star\rq{}s stellar winds. The observed reddening of starlight that has been noticed by astronomers since the nineteenth century is well known to be due to fine dust material distributed in the space between stars and to the observer. Dust makes up about 1$\%$ of the mass of interstellar matter and also plays an important role in the different stages of stellar evolution (\citet{Apai2010}). It is believed that dust has to be distributed into the interstellar medium by being propelled through stellar winds but a complete picture of the motion of dust in the Galaxy yet remains to be understood. The almost ubiquitous nature of dust that fills the Galaxy may be aided by the galactic gravitational force ${\bf F_G}$. This newly discovered dynamic should be investigated further since it may have a profound effect on stellar evolution that has yet to be recognized. 
\medskip\medskip\medskip

\noindent
{\em Future Missions Leaving the Solar System }
\medskip\medskip\medskip

Our own Sun generates a magnetic field that propagates outward in all directions carried by the solar wind, forming a magnetic bubble around the planets of the solar system  called the heliosphere. Within the last decade we have probed the outer reaches of the heliosphere with the Voyager 1 and 2 spacecraft until they passed through the heliosphere boundary or heliopause at about 120 AU and into the interstellar medium (ISM). Current models suggest that the heliosphere is a bubble which may also be more drawn out downstream of the stellar wind than upstream. NASA is studying a new mission to explore the outer heliosphere and ISM called Interstellar Probe or IP (\citet{Brandt2019}, \citet{ McNutt2019}). IP is the first mission designed and instrumented specifically to study both the outer heliosphere and the near and distant ISM. 

Science targets for the IP include a flyby of a selected KBO, the physics of the ISM, and the first external images of the Extragalactic Background Light (EBL) beyond the Zodiacal cloud and image our own heliosphere. The science payload is expected to include particle and fields detectors, a dust detector, along with optical and infrared imaging cameras. IP is expected to have a nominal design lifetime of 50 years. IP is being designed to travel more than twice the distance of the Voyager spacecraft, out to about 400 AU with a goal target of about 1000 AU. The IP is being designed to have speeds ranging from 8 to 15 AU/year. The IP will be humanities\rq{} first step in truly interstellar exploration and the first mission that will have the opportunity to detect ${\bf F_{G,S}}$. To maximize this force on the spacecraft  providing additional acceleration to the mission our recommendation would be to target the mission to fly in the general direction of the Galactic Centre. This would also have the advantage of flying through the heliopause and heliosheath on the unexplored flanks of the heliosphere and provides an opportunity to obtain an image cross-section determining its entire shape which can then be more appropriately compared to the model calculations. 

 In addition, IP could be our first opportunity to study the very edge of the transition region where $F_{G,S}$ could be measured by releasing, at an appropriate distance from the Sun,  a small unpowered secondary payload whose displacement would be unaffected by the radioisotope power and thermal sources of IP and could be precisely measured.
\medskip

It is remarked that when measuring the effect of $F_{G,S}$ on the displacement of a spacecraft near to the Sun, say 50 AU distant, there are other accelerations acting on the spacecraft. For example, the gravitational field of the Alpha Centauri system would impart an acceleration that could be second order but could also be
of a similar in magnitude as $F_{G,S}$, that the detection process would have to consider. The study of this is beyond the scope of this paper.

\section{Conclusions} \label{sec:Conclusions}
\medskip

The results presented here demonstrate the existence of a resultant gravitational force of the Galaxy, largely due to the mass of the halo, dominated by dark matter. It may be important to model for spacecraft moving far from the Sun on long duration missions where this force can build up. It may be possible to detect on a proposed mission called Interstellar Probe.  The existence of this force has many implications for planetary astronomy and astrophysics, as pertains to the motion of objects in the distant solar system and outside our solar system. 
\medskip\medskip 


\section*{Acknowledgements}

\noindent
  E.B. would like to thank Marian Gidea for comments on improving an estimate and Urs Frauenfelder\rq{}s comments, and he is very grateful for several comments by Michael Strauss and Scott Tremaine which improved this paper.  E.B. also thanks  David Spergel helpful comments and Neil de Grasse Tyson for discussions  in 2007. Thanks also to Frans Pretorius.    J.G. gratefully acknowledges discussions with R. Roncoli (Voyager spacecraft), H. Winters and C. Hersman (New Horizons) and with M. Brown on the status of observing Planet 9.  Funding:  E.B. was partially funded by NSF grant DMS-1814543.  (Author Contributions: E.B. contributed  the initial concept and modeling.   J.G. contributed implications to spacecraft, astronomy and astrophysics.)
\medskip\medskip

{\it Data Availability}
\medskip

There are no new data associated with this article


\bibliographystyle{mnras}

\begin{thebibliography}{99}


\bibitem[\protect\citeauthoryear{AHearn}{2006}]{AHearn2006}
A\rq{}Hearn MF,  2006
Whence comets?
{\it Science} {\bf 314} 1708-1709



\bibitem[\protect\citeauthoryear{Aghanim}{2019}]{Aghanim2020}
Aghanim N, et.al,  2020
Planck 210 results. I. Overview and the cosmological legacy of Planck
{\it A\&A} {\bf 641} A1


\bibitem[\protect\citeauthoryear{Apai}{2010}]{Apai2010}
Apai D, Lauretta DS,2010
{\it Protoplanetary Dust: Astrophysics and Cosmochemical Perspectives}
(Cambridge: Cambridge University Press) 

\bibitem[\protect\citeauthoryear{Arbey}{2021}]{Arbey2021}
Arbey A, Mahmoudi F,  2021,
Dark matter and the early universe: a review
{\it Progress in Particle and Nuclear Physics} {\bf 119} 103865

\bibitem[\protect\citeauthoryear{Bailer-Jones}{2018}]{Bailer-Jones2018}
Bailer-Jones CAL, Farnocchia D, Meech K,  et. al., 2018
Plausible home stars of the interstellar object: \lq{}Oumuamua found in Gaia DR2
{\it AJ} {\bf 156} 295

\bibitem[\protect\citeauthoryear{Batygin}{2016}]{Batygin2016}
Batygin K, Brown ME,  2016
Evidence for a giant planet in the solar system 
{\it AJ} {\bf 151} 22 

\bibitem[\protect\citeauthoryear{Batygin}{2019}]{Batygin2019}
Batygin K, Adams KF, Brown ME, Becker J, 2019
The planet nine hypothesis {\it Phys. Reps.} {\bf 805} 1-53

\bibitem[\protect\citeauthoryear{Belbruno}{2004}]{Belbruno2004}
 Belbruno E, 2004
  {\it Capture Dynamics and Chaotic Motions in Celestial Mechanics}
  (Princeton: Princeton University Press)
  
  
 \bibitem[\protect\citeauthoryear{Belbruno}{2010}]{Belbruno2010}
 Belbruno E, Gidea M, Topputo F,  2010
 Weak stability boundary and invariant manifolds
 {\it SIAM J Appl. Dyn. Sys.} {\bf 9} 1061-1089  
   
 \bibitem[\protect\citeauthoryear{Bennett}{2003}]{Bennett2003}
Bennett C, {\it et. al.},  2019
The Microwave Anisotropy Probe (WMAP) mission {\it ApJ} {\bf 583} 1-23        

\bibitem[\protect\citeauthoryear{Bovy}{2014}]{Bovy2014}
 Bovy J,  2014
 galpy: A python library for galactic dynamics
{ApJ S}, {\bf{216} }  29

\bibitem[\protect\citeauthoryear{Brandt}{2019}]{Brandt2019}  
Brandt  PC,  McNutt R L ,  Paul MV,   Lisse CM,  et. al.,  2019
Humanities first explicit step in reaching another star: the interstellar probe mission
{\it JBIS} {\bf 72} 202-212

\bibitem[\protect\citeauthoryear{Callingham}{2019}]{Callingham2019}
Callingham TM, Cautun M, Deason  AJ , Frenk CS, et. al.,  2019
The mass of the galaxy from satellite dynamics
{\it MNRAS}  {\bf 484} 5453-5467



 \bibitem[\protect\citeauthoryear{Dillamore}{2021}]{Dillamore2021}
Dillamore A, Belokurov V, Font, AS, McCarthy IG, 2021
Merger-induced galaxy transformations in the ARTEMIS simulations
{\it MNRAS}  (in press)



\bibitem[\protect\citeauthoryear{Eilers}{2019}]{Eilers2019}
  Eilers A-C, Hogg, DW, Rix, H-W, Ness, KN, 2019
  The circular velocity curve of the Milky Way from 5 to 25 kpc
  {\it ApJ} {\bf 871} 120
  
\bibitem[\protect\citeauthoryear{Guerra}{2017}]{Guerra2017}
Guerra A, Francisco F, Gil P, Bertolami O,  2017
Estimating then thermally induced acceleration of the New Horizons spacecraft
{\it Phys. Rev. D} {\bf 95} 124027

 \bibitem[\protect\citeauthoryear{Helmi}{2008}]{Helmi2008}
Helmi A , 2008 
 The stellar halo of the galaxy
{\it A\&A Rev.} {\bf 15} 145
   
     \bibitem[\protect\citeauthoryear{Heisler}{1986}]{Heisler1986}
 Heisler J, Tremaine S,  1986
 The influence of the galactic tidal field on the oort comet cloud
 {\it Icarus} {\bf 65} 13-26

\bibitem[\protect\citeauthoryear{Hernquist}{1990}]{Hernquist1990}
Hernquist L, 1990
An analytical model for spherical galaxies and bulges
{\it ApJ} {\bf 409} 548 

\bibitem[\protect\citeauthoryear{Licquia}{2013}]{Licquia2013}
Licquia, T, Newman, J. 2013
Total stellar mass, color and luminosity of the Milky Way
{American Astronomical Society, AAS Meeting no. 221} , id.254.11 


\bibitem[\protect\citeauthoryear{McNutt}{2019}]{McNutt2019}
McNutt, R L,  Wimmer-Schweingruber RF, Gruntman M,   Krimigis SM,  et. al.,   2019
Near-term interstellar probe: first step
{\it Acta Astron.} {\bf 162} 284
 
 
\bibitem[\protect\citeauthoryear{Meech}{2017}]{Meech2017}
Meech K, Weryk R, Micheli M, {\it et. al.}, 2017
A Brief visit from a red and extremely elongated interstellar asteroid
{\it Nature} {\bf 552}, 378-381
 
\bibitem[\protect\citeauthoryear{Mewhinney}{2003}]{Mewhinney2003}
Mewhinney M,  2003
Pioneer 10 spacecraft sends last signal
{\it NASA Release: 03-082HQ}

\bibitem[\protect\citeauthoryear{Micheli}{2018}]{Micheli2018}
Micheli M, Farnochia  D, Meech K,  {\it  et. al.},  2018
Non-gravitational acceleration in the trajectory of 1I/2017 U1 (\lq{}Oumuamua) 
{\it Nature} {\bf 559} 223-226


\bibitem[\protect\citeauthoryear{Navarro}{1996}]{Navarro1996}
Navarro JF,  Frenk CS, White SDM, 1996
The structure of cold dark matter halos
{\it ApJ } {\bf 462} 563

 \bibitem[\protect\citeauthoryear{Piffl}{2014}]{Piffl2014}
Piffl T, {\it et. al.}, 2014
Constraining the galaxy\rq{}s dark halo with RAVE stars
{\it MNRAS}  {\bf 445  } 3133-3151


\bibitem[\protect\citeauthoryear{Racca}{2002}]{Racca2002}
Racca GD,   {\it et. al.},  2002
SMART-1 mission description and development status
{\it Planetary and Space Science} {\bf 50}  1323-1337  



\bibitem[\protect\citeauthoryear{Roncoli}{2012}]{Roncoli2012}
Roncoli R,  Fujii K 2012
Mission design overview for the Gravity Recovery and Interior Laboratory (GRAIL) Mission
{\it Proceedings AIAA Guidance, Navigation, and Control Conf. } Paper AIAA 2010-8383 


\bibitem[\protect\citeauthoryear{Rogers}{2014}]{Rogers2014}
Rogers G, Flanagan S, Stanbridge D,  2014
Effects of radioisotope thermoelectric generator on dynamics of the New Horizons Spacecraft
{\it Proceedings AAS}, Paper AAS-14-122  


\bibitem[\protect\citeauthoryear{Spergel}{2003}]{Spergel2003}
Spergel DN,  2003
First-Year Wilkinson Microwave Anisotropy Probe(WMAP) observations: determination of cosmological
parameters
{\it ApJ S} {\bf 148} 175-194 


\bibitem[\protect\citeauthoryear{Topputo}{2009}]{Topputo2009}
 Topputo F,  Belbruno E,   2009
Computation of weak stability boundaries: Sun-Jupiter 
 {\it Cel. Mech. Dyn. Astr.} {\bf 105} 3-17 


\bibitem[\protect\citeauthoryear{Turyshev}{2010}]{Turyshev2010}
 Turyshev SG, Toth VT,  2010
The Pioneer anomaly
 {\it Living Rev. Relativity} {\bf 13} 4 
 
  \bibitem[\protect\citeauthoryear{Turyshev}{2011}]{Turyshev2011}
  Turyshev SG, Toth  VT, Ellis J, Markwardt C, 2011 
  Support for temporally varying behavior of the Pioneer anomaly from the extended Pioneer 10 and 11 doppler data sets.
{\it   Phys. Rev. Letters} {\bf 107}  8 

   \bibitem[\protect\citeauthoryear{Turyshev}{2012}]{Turyshev2012}
  Turyshev SG, Toth VT, Kinsella G, Lee S-C, Lok SM, Ellis J, 2012
 Support for the thermal origin of the Pioneer anomaly
{\it Phys. Rev. Letters} {\bf 108} 24 

\bibitem[\protect\citeauthoryear{Watkins}{2019}]{Watkins2019}
Watkins L, van der Markel RP, Sohn ST , Evans NW,  2019
Evidence for an intermediate-mass Milky Way from {\it Gaia}  DR2 halo globular cluster motions
{\it ApJ} {\bf 873} 118        

\bibitem[\protect\citeauthoryear{Wechsler}{2018}]{Wechsler2018}
Wechsler R H, Tinker J L, 2018
The connection between galaxies and their matter halos
{\it Annu. Rev. Astron. Astrophys.} {\bf 56} 435-487

 











\end{thebibliography}

\end{document}